\documentclass{article}

\def\bSig\mathbf{\Sigma}

\newcommand{\logit}{\text{logit}}
\usepackage{amsmath}
\usepackage{booktabs}
\usepackage{multicol}
\usepackage{enumitem}
\usepackage[font=small]{caption}
\usepackage{graphicx}
\usepackage{arxiv}
\usepackage[utf8]{inputenc} % allow utf-8 input
\usepackage[T1]{fontenc}    % use 8-bit T1 fonts
\usepackage{hyperref}       % hyperlinks
\usepackage{url}            % simple URL typesetting
\usepackage{booktabs}       % professional-quality tables
\usepackage{amsfonts}       % blackboard math symbols
\usepackage{nicefrac}       % compact symbols for 1/2, etc.
\usepackage{microtype}      % microtypography
\usepackage{lipsum}
\usepackage{comment}
\usepackage{graphicx}
\graphicspath{ {./images/} }
\usepackage{subcaption}
\usepackage{floatrow}
\usepackage{sidecap}
\sidecaptionvpos{figure}{c}
\usepackage{graphicx}
\newtheorem{definition}{Definition}
\usepackage{lmodern}
\usepackage{doi}
\usepackage{algpseudocode}
\usepackage{sidecap}
\usepackage{multicol}
\usepackage{amsmath}
\usepackage{multirow}
\newcommand{\E}{\mathbb{E}}

\newcommand{\R}{\mathbb{R}}

\usepackage{natbib}
\usepackage{color}
\usepackage[ruled]{algorithm2e}
\usepackage{algpseudocode}
\usepackage[dvipsnames]{xcolor}
\usepackage{indentfirst}
\usepackage{listings}
\usepackage{xcolor}
\newtheorem{theorem}{Theorem}
\RequirePackage{cite}
\setcitestyle{numbers,square}
\definecolor{Amethyst}{rgb}{0.15, 0.38, 0.61}
\definecolor{pastelgray}{rgb}{0.81, 0.81, 0.77}
\definecolor{babyblue}{rgb}{0.68, 0.78, 0.81}
\definecolor{mauvetaupe}{rgb}{0.57, 0.37, 0.43}

\lstdefinestyle{mystyle}{
    commentstyle=\color{Amethyst},
    keywordstyle=\color{mauvetaupe},
    stringstyle=\color{babyblue},
    breakatwhitespace=false,         
    breaklines=true,                 
    captionpos=b,                    
    kepdfpaces=true,                 
    numbers=none,    
    columns=flexible,
    basicstyle={\small\ttfamily},
    showspaces=false,                
    showstringspaces=false,
    showtabs=false,                  
    tabsize=2
}

\title{Signature-Informed Selection Detection: \\ A Novel Method for Multi-Locus Temporal Population Genetic Model with Recombination}
\author{
 Ritabrata Dutta\\
  Department of Statistics\\
  University of Warwick\\
  Coventry CV4 7AL, U.K.\\
  \texttt{ritabrata.dutta@warwick.ac.uk} \\
  \And  
 Yuehao Xu \\
  Department of Applied Statistics\\
  Johannes Kepler University Linz\\ 
  Linz 4040, Austria\\
  \texttt{yuehao.xu@jku.at} \\
  \And
 Sherman Khoo\\
  School of Mathematics\\
  University of Bristol\\
  Bristol BS8 1UG, U.K.\\
  \texttt{sherman.khoo@bristol.ac.uk} \\
   \And
 Francesca Basini\\
  Department of Statistics\\
  University of Warwick\\
  Coventry CV4 7AL, U.K.\\
  \texttt{Francesca.Basini.1@warwick.ac.uk} \\
  \And  
 Andreas Futschik \\
  Department of Applied Statistics\\
  Johannes Kepler University Linz\\
  Linz 4040, Austria\\
  \texttt{andreas.futschik@jku.at} \\
}

\begin{document}
\maketitle
\noindent

\begin{abstract}
In population genetics, there is often interest in inferring selection coefficients. This task becomes more challenging if multiple linked selected loci are considered simultaneously.
 For such a situation, we propose a novel generalized Bayesian framework where we compute a scoring rule posterior for the selection coefficients in multi-locus temporal population genetics models. As we consider trajectories of allele frequencies over time as our data, we choose to use a signature kernel scoring rule - a kernel scoring rule defined for high-dimensional time-series data using iterated path integrals of a path (called signatures). We can compute an unbiased estimate of the signature kernel score using model simulations. This enables us to sample asymptotically from the signature kernel scoring rule posterior of the selection coefficients using pseudo-marginal MCMC-type algorithms. Through a simulation study, we were able to show the inferential efficacy of our method compared to existing benchmark methods for two and three selected locus scenarios under the standard Wright-Fisher model with recombination and selection. We also consider a negative frequency-dependent selection model for one and two locus scenarios, and also joint inference of selection coefficients and initial haplotype frequencies under 
 the standard Wright-Fisher model. Finally, we illustrate the application of our inferential method for two real-life dataset. More specifically, we consider a data set on Yeast, as well as data from an Evolve and Resequence (E\&R) experiment on {\em Drosophila simulans}.
\end{abstract}

\section{Introduction}
\label{sec:Intro}
Natural selection is a key concept in evolution. It helps advantageous traits spread through a population while limiting the spread of harmful ones. Selective pressure changes the genetic composition of populations over time, as advantageous alleles become more prevalent and deleterious alleles are eliminated. Natural selection significantly influences how species adapt to their surroundings. By affecting genetic diversity within and across populations, it contributes
to the genetic structure of a species.  

Several statistical methods have been proposed to infer selection, including both estimates and hypothesis tests. We refer to  \citep{fay2000hitchhiking}, \citep{kim2002detecting}, \citep{pavlidis2013sweed}, \citep{foll2008genome}, \citep{kim2004linkage}, \citep{nielsen2005genomic}, \citep{ferrer2014detecting}, among others.  \citep{vitti2013detecting} 
and \citep{booker2017detecting}
offer reviews on the detection of selection within the broader scope of evolutionary genetics.

Our focus will be on signals of selection in temporal genetic data. Methods that use 
allele frequencies at multiple time points
include \citep{barata2023bait}, a Bayesian approach that employs the Moran model as a foundational structure for estimating selection coefficients specifically within the context of Evolve and Resequence (E\&R) experiments. The Composite-Likelihood Approach for Evolve and Resequence Experiments (CLEAR), developed by \citep{iranmehr2017clear} aims to detect selected loci and to infer their selection coefficients. A Bayesian approach is the Wright-Fisher ABC (WFABC) method, introduced by \citep{foll2015wfabc}. 
These methods operate at a single locus level. 

In contrast,
the method developed by \citep{he2020detecting} considers two linked loci.  
Specifically, it employs a Hidden Markov Model (HMM) coupled with a two-locus Wright-Fisher diffusion model (called HMM-WFD) that includes selection and recombination. This integration allows for the consideration of recombination. For the estimation of posterior distributions of the selection coefficients, the authors utilize the particle marginal Metropolis-Hastings algorithm. Further work on inferring selection based on time series data
includes \citep{schraiber2016bayesian}, \citep{malaspinas2012estimating},\citep{steinrucken2014novel},\citep{bollback2008estimation},\citep{lacerda2014population}, \citep{paris2019inference},\citep{feder2014identifying} and \citep{terhorst2015multi}.

Our study centers on temporal allele and haplotype frequency data. We interpret an allele as one of two variants of a SNP denoted as $A_1$ and $A_2$. A haplotype consists of $\ell$ such SNP loci. There are $2^\ell$ possible different haplotypes with $\ell$ loci. Observing the haplotype frequencies over time provides information on whether natural selection acts on one or more of the underlying SNPs. In cases of linked loci, this dynamic gains further complexity. Selection's impact on one locus extends to its linked loci, influencing their allele frequencies. Phenomena like genetic hitchhiking or background selection play a role. Our interest lies in disentangling selective forces within a multi-locus framework. We consider a Wright-Fisher model with recombination where each SNP may have a separate fitness effect. Inferring selection coefficients in a multi-locus system poses additional challenges compared to single-locus inference but makes it possible to disentangle signals from multiple SNPs in the presence of linkage. With the recent interest in polygenic adaptation,  there is a clear need for a method capable of accurately estimating the selection coefficients of multiple loci; our study aims to tackle this challenge. 

The majority of current approaches in the field use approximation techniques, e.g., approximate Bayesian computation (ABC) \citep{lintusaari2017fundamentals} because exact likelihoods are usually unavailable in more complex population genetic situations. In this study, we present a novel method for approximating the likelihood through signature kernel score \citep{kiraly2019kernels}. 
Originating from rough path theory \citep{lyons1998differential}, signatures have proven useful in machine learning applications \citep{graham2013sparse, yang2022developing, kidger2019deep}, which can encode information about a path, constructed by iteratively integrating the path concerning time and taking the tensor product of the resulting iterated integrals. Signatures effectively capture patterns in sequential data, such as temporal haplotype frequency data. For an introduction and survey on applications, see \citep{chevyrev2016primer}. We employ the signature kernel score \citep{kiraly2019kernels, salvi2021signature} within the framework for generalized Bayesian likelihood-free inference using scoring rules as in \citep{pacchiardi2023generalized} to get an approximation of the likelihood (and hence an approximation to the posterior, called scoring-rule posterior \citep{giummole2019objective}). Given an unbiased estimate of the proposed scoring rule can be constructed using simulations from the Wright-Fisher model, we use a pseudo marginal Markov chain Monte Carlo (MCMC) \citep{Andrieu_2009} type algorithm, to generate posterior samples of the parameters of interest using the Metropolis-Hastings algorithm \citep{hastings1970monte, pacchiardi2023generalized}.

We investigate two-locus and three-locus models with recombination within the standard Wright-Fisher framework and in a frequency-dependent selection regime. To evaluate the performance of our proposed method, we compare it using simulations with the methods proposed by \citep{he2020detecting} and \citep{taus2017quantifying} as a benchmark. These methods were selected as benchmarks because they operate within a similar underlying model framework of temporal allele frequency data. Despite these methodological commonalities, they employ different modelling techniques and inferential approaches compared to ours. 

In Section~\ref{sec:models}, we introduce the discrete-time Wright-Fisher model (Subsection~\ref{subsection:WFD}) and a frequency-dependent selection model (Subsection~\ref{sec:FDS}). 
After motivating our proposed method in contrast to other likelihood-free inference methods in section  \ref{sec:methods}, we describe the scoring rule posterior and how to efficiently sample from it in Subsection~\ref{sec:scoringrulep} and \ref{sec:sampling}. The signature kernel score used in this work is described in Subsection~\ref{sec:signatures}, starting with the necessary data pre-processing. 
In Section~\ref{sec:inf_selcof}, we illustrate how our methodology can be used to infer selection coefficients for the Wright-Fisher model (Section~\ref{sec:Inf_WF}) and a negative frequency-dependent selection model (Section~\ref{sec:Inf_NFDS}). For benchmarking, we also compare our approach with other methods. In Section~\ref{sec:yeast_data}, we consider a real data set containing temporal allele frequencies collected from Yeast \citep{phillips2021}. Then in Section~\ref{sec:joint_selcof_inthapfreq}, we extend our methodology to jointly infer selection coefficients and initial haplotype frequencies.
We consider both simulated allele frequency data under a two-locus scenario, and experimentally observed allele frequency data from {\em Drosophila simulans} for a three-locus scenario \citep{barghi2019genetic}, assuming the Wright-Fisher model. 
In the concluding Section \ref{section:conclusion}, we summarize 
our research findings and discuss where and when our method could be useful.

%\section{Method and Material}

\section{Temporal Population Genetic Models}
\label{sec:models}

\label{section:wfmodels}
Since our proposed method is fundamentally simulation-based, we will use a simulator capable of generating data that mimics real-world observations, enabling comparison between the simulated and observed data sets. We would consider a Wright-Fisher model with selection and a negative frequency-dependent selection model \citep{svensson2019frequency} in this paper. 

The simulations for these models could be carried out either in an exact discrete-time setup or in continuous time. To simulate in continuous time,
the discrete-time Wright-Fisher model
is approximated by a stochastic differential equation (Wright-Fisher diffusion).
While the Wright-Fisher diffusion, particularly with the Euler-Maruyama (EM) method \citep{platen2010numerical}, might be faster for data sets spanning over many generations, it faces inherent boundary issues \citep{dangerfield2012boundary}.
Such issues can cause the numerical solution to exceed the [0, 1] boundary, leading to practical problems. A recent contribution to continuous time simulation is \citep{sant2023ewf}. It introduces an interesting new exact approach, but requires substantial computational efforts.

Given our focus on precision, computational efficiency, and minimal bias, we have chosen the discrete-time Wright-Fisher model for our investigations. 
Existing software for forward-time simulation of such a model includes SLiM3 \citep{haller2019slim}, MimiCREE2 \citep{vlachos2018mimicree2}, fwdpp \citep{thornton2014c++}, and simuPOP \citep{peng2005simupop}. To ensure efficiency in combination with our methodological code, we coded our own simulator as explained in Section~\ref{sec:models} below.

\subsection{Discrete Time Wright-Fisher Model}
\label{subsection:WFD}
We consider a discrete-time Wright-Fisher model involving $\ell$ loci
with recombination. More specifically, we focus on values of $\ell$ between one and three. In the three-locus model, the loci are denoted by $A$, $B$,  and $C$, with $B$ being between $A$ and $C$ on the chromosome. The per-generation recombination rates $r_{AB}$ and $r_{BC}$ provide the recombination probabilities between neighboring loci.
Since we assume that there are two alleles at each locus, an $\ell$-locus model translates
into a maximum of $2^\ell$ haplotypes that represent all possible allelic combinations at the loci.

We aim to infer selection parameters for the loci. Our main focus will be on the basic multiplicative viability fitness model with selection coefficients $\theta=(s_A,s_B,s_C)$ in the three locus case. In principle, alternative selection models can also be inferred using our approach. In this case, further model parameters such as dominance coefficients, the effective population size $N_e,$ and epistasis, may be introduced. The accuracy of inference in such more complex models will depend on how much signal the data provides concerning additional parameters. To illustrate the extension to an alternative model, we consider frequency-dependent selection in subsection~\ref{sec:FDS} below.

For our multiplicative viability selection model,
a detailed explanation of how to obtain diploid fitness 
coefficients $w_{ij}$
for an individual carrying the haplotypes $i$ and $j$ can be found in \citep{burger2000mathematical}. 
With $h_{t,i}$ being the frequency of haplotype $i$ at time $t$,
we simulate the frequencies of haplotypes in the next generation $t+1$
following \citep{burger2000mathematical}:
\begin{equation}
\begin{aligned}
&\rho_{ij}(t) =  \sum_I r_I (h_{t,i} h_{t,j} - h_{t,i_{I}j_{J}}h_{t,j_{I}i_{J}}), \ \tilde{h}_{t+1,i} =\frac{\sum_{j} w_{ij}
(h_{t,i} h_{t,j} - \rho_{ij}(t))
}{\sum_{i}\sum_{j}w_{ij}
h_{t,i} h_{t,j}}  \\
&\left(h_{t+1,1},\ldots,h_{t+1,i},\ldots,h_{t+1,2^\ell}\right) \sim \frac{1}{2N} \times\text{Multinomial}(2N, (\tilde{h}_{t+1,1},\ldots,\tilde{h}_{t+1,2^\ell}))
\label{eq:3locus}
\end{aligned}
\end{equation}
In Equation (\ref{eq:3locus}), the term $ \tilde{h}_{t+1,i} $ denotes the expected frequency of the $i$-th haplotype at time $ t + 1$, 
conditional on the fitness term $w_{ij}$ and the frequencies at time $t$.  The subsequent stochastic step, embodied by multinomial sampling, introduces randomness reflecting genetic drift. 
The terms $\rho_{ij}$ account for 
recombination between the haplotypes, where $r_I$ are the corresponding recombination rates. Equation (\ref{eq:3locus}) can be adapted to any number of loci and corresponding haplotypes. Further, the initial haplotype frequency $h_{t_0, 1:2^l}$ would be assumed to be known in Section~\ref{sec:inf_selcof} and in Section~\ref{sec:joint_selcof_inthapfreq} we would jointly infer the initial haplotype frequencies and the selection coefficients.

The output of our simulator will be a matrix 
$\mathbf{x}=\lbrace a_{t,i}: t=t_0,\ldots, T, i=1,\ldots,l \rbrace$
where columns represent the selected loci, and rows the time points at which data are available. The entries are allele frequencies of the considered selected loci, resulting from the underlying haplotype frequencies $\lbrace h_{t,i}: t=t_0,\ldots, T, i=1,\ldots,2^\ell \rbrace$. They are provided from the start time $t_0$ to $T =t_0+K\times\Delta t$, at intervals of length  $\Delta t$. We thus assume that haplotype frequencies are observed at $K+1$ equidistant time points. This assumption is not necessary for our method to work as long as the simulated data are available at the same times as the observed data. 

We will use the symbol $\mathbf{x}^{\theta}$ for data simulated using parameter $\theta$. Our parameter $\theta$ consists of the unknown selection coefficients.
We use independent Uniform[-1.0,1.0] distributions as their prior, containing all plausible parameter values in an actual application. 

\subsection{Negative Frequency Dependent Selection \label{sec:FDS}}

A common alternative selection scenario is negative frequency-dependent selection, where the fitness of a haplotype decreases with its frequency. A well-known example
of negative frequency-dependent selection is, for instance, self-incompatibility in plants. Here we would consider such a model for allele frequency dynamics proposed for a single-locus system as in \citep{svensson2019frequency} (p.\ 1246). The only difference from the previously considered Wright-Fisher model is in the way we calculate the fitness coefficients $w_{ij}$, in which we would replace the selection coefficients by selection coefficients multiplied by the allele frequency of the corresponding loci at that time point. Hence, in this case, the fitness coefficient $w_{ij}$ would be dependent on time via the temporal changes in allele frequencies. 

\section{Generalized Bayesian likelihood-free inference}
\label{sec:methods}
As in many population genetic models, the likelihood function is analytically intractable for the multivariate temporal population genetic models considered above.
Nevertheless, the considered models permit direct sampling, a distinctive feature that we capitalize upon. This capability to derive samples directly from the model suggests using a likelihood-free inference framework. This approach entails generating data by running simulations based on the model parameters. The samples we generate from the simulator contain a spectrum of potential outcomes under the model, given the population parameters. 
By using methods available under the broad umbrella of likelihood-free inference (LFI), these samples let us make inferences without having to confront the mathematical intractability of the likelihood function \citep{beaumont2002approximate, foll2015wfabc, collin2021extending}. Two of the most popular classes of LFI methods are approximate Bayesian computation (ABC) \citep{lintusaari2017fundamentals} and Bayesian synthetic likelihood \citep{price2018bayesian}. They approximate the intractable likelihood function either implicitly or explicitly using these samples. The asymptotic contraction of these approximate posteriors towards the true parameter value depends upon the choice of the summary statistics and some conditions being satisfied by the chosen summary statistics \citep{frazier2018asymptotic, li2018asymptotic, frazier2023bayesian}, which are difficult to verify in practice.

In this work, we choose a different likelihood approximation method, the scoring rule posterior framework of \citep{pacchiardi2023generalized}, which facilitates likelihood-free inference using a suitable scoring rule for a given type of data. This is similar to Bayesian synthetic likelihood in the sense that it provides an explicit approximation to the likelihood. However, it is easier to verify that the scoring rule posterior contracts to the true parameter value when a strictly proper scoring rule is chosen \citep{pacchiardi2023generalized}. In this work, we choose a strictly proper scoring rule, the signature kernel score proposed in \citep{kiraly2019kernels, salvi2021signature}, that is particularly effective for our time series of haplotype frequency data. The signature kernel is defined via an inner product on path signatures \citep{chevyrev2016primer}, which are infinite-dimensional summary statistics defined via iterated path integrals and known to be able to extract higher-order information from time-series data. 

In \citep{dyer2022amortised} and \citep{dyer2023approximate}, the path signatures were used as summary statistics in an ABC framework.
Further \citep{issa2023nonadversarial} used the signature kernel with a kernel scoring rule 
in the context of training a generative time-series model defined by a neural stochastic differential equation. 
In contrast to these works, we use the signature kernel scoring rule to define a scoring rule posterior as part of the generalized Bayesian likelihood-free inference (GBLFI) framework \citep{pacchiardi2023generalized}.

\subsection{Scoring rule posterior}
\label{sec:scoringrulep}
In this subsection, we detail the usage of the scoring rule posterior for the generalized Bayesian likelihood-free inference setup. 

If we had access to the likelihood function $p(\mathbf{x}^{obs} \mid \theta)$ for our model with
observed data $\mathbf{x}^{obs}$,
parameters $\theta$, and a prior distribution $p(\theta)$ on our parameters, then the posterior distribution $\pi(\theta \mid \mathbf{x}^{obs})$ could be obtained using Bayes Theorem as follows,  
$$
\begin{aligned}
\pi(\theta \mid \mathbf{x}^{obs}) \propto p(\theta) p( \mathbf{x}^{obs}\mid \theta) =  p(\theta) \exp \left\{ \log p(\mathbf{x}^{obs} \mid \theta) \right\}.
\end{aligned}
$$
Due to the aforementioned intractability of the likelihood function $p(\mathbf{x}^{obs} \mid \theta)$, we are unable to evaluate this function, and thus will not be able to proceed with obtaining the posterior distribution.
As in \citep{gneiting2007strictly},
we consider loss functions $S(P_{\theta},\mathbf{x})$ known as scoring rules that measure the fit between the distribution $P_{\theta}$ of the data under parameter $\theta$, and 
an observed data point $\mathbf{x}$. See equation (\ref{eq:unbiased_kernel_score}) for
the scoring rule we are focusing on.

The \textit{scoring rule posterior} is then defined as follows:
\begin{equation}
\label{eq:SR_post}
\pi_{S}\left(\theta \mid \mathbf{x}^{obs}\right) \propto p(\theta) \exp \left\{-w S\left(P_{\theta}, \mathbf{x}^{obs}\right)\right\}
\end{equation}

Comparing the two expressions, we note that the (negative) log-likelihood function can itself be considered a scoring rule (known as log-score \citep{Dawid_2014}) and that we have introduced an additional parameter $w$, which is known as the learning rate in generalized Bayesian inference \citep{holmes2017assigning} controlling the relative weighting of the observations relative to the prior. For this study, we try to keep $w$ fixed at $1$, if not otherwise specified. $w$ could be tuned to adapt the contraction rate of the scoring rule posterior \citep{pacchiardi2023generalized}. 

A scoring rule is said to be \textit{proper}, if its expected value with respect to the observed data $ \mathbf{x}^{obs}$ is minimized when the assumed distribution $P_\theta$ is equal to the
distribution $P_0$ generating the observed data $\mathbf{x}^{obs}$. If $P_\theta = P_0$ is the unique minimum, the scoring rule is said to be \textit{strictly proper}. A strictly proper scoring rule is desired as this property further ensures that the best prediction one can make, according to the scoring rule, is exactly the truth. The existence of a unique minimum, i.e., the true distribution, provides a strong motivation for models to accurately capture the underlying distribution of the data. Theoretical properties of asymptotic normality and generalization bound for scoring rule posteriors when the scoring rule is strictly proper have been studied in \citep{giummole2019objective, pacchiardi2023generalized}. Additionally, the scoring rule posterior for some scoring rules (e.g., energy score or kernel score) exhibits robustness against outliers compared to the standard Bayes posterior using log-score \citep{pacchiardi2023generalized}. Further to note that the signature kernel score used in this work is strictly proper \citep{salvi2021signature} and also robust, being a kernel scoring rule \citep{pacchiardi2023generalized}.

\subsection{Sampling the scoring rule posterior}
\label{sec:sampling}
While direct sampling from the Scoring Rule posterior is still infeasible so far, \citep{pacchiardi2023generalized} notes that approximate sampling from a scoring rule posterior in the likelihood-free framework is feasible when there exists an unbiased estimate of the scoring rule w.r.t. $P_{\theta}$. If there exists an unbiased estimate $ \widehat S(\mathcal{X}^{\theta,m}, \mathbf{x}^{obs}) $ of $S(P_{\theta}, \mathbf{x}^{obs})$ using $m$ i.i.d.\ simulations $\mathcal{X}^{\theta,m}=\lbrace \mathbf{x}^{\theta,1},\ldots, \mathbf{x}^{\theta,m}\rbrace$ from the distribution $P_{\theta}$, we can get an estimate of the scoring rule posterior \begin{equation}
\pi_{\widehat{S}}^{(m)}(\theta \mid \mathbf{x}^{obs}) {\propto} \pi(\theta)\left[\exp\{ - w \widehat S(\mathcal{X}^{\theta, m}, \mathbf{x}^{obs}) \} \right].
\end{equation}

We also need to emphasize that the revised target distribution $\pi_{\widehat{S}}^{(m)}(\theta \mid \mathbf{x}^{obs})$ is not an unbiased estimate of $\pi_{S}\left(\theta \mid \mathbf{x}^{obs}\right)$. Nonetheless, when the number of simulations $m$ tends towards infinity, there is the following result from \citep{pacchiardi2023generalized}:
\begin{theorem}\label{Th:auxiliary_lik} 
If $\widehat{S}\left(\mathcal{X}^{\theta,m}, \mathbf{x}^{obs}\right)$ converges in probability to $S\left(P_\theta, \mathbf{x}^{obs}\right)$ as $m \rightarrow \infty$, 
then under some minor technical assumptions:
$$
\lim _{m \rightarrow \infty} \pi_{\widehat{S}}^{(m)}(\theta \mid \mathbf{x}^{obs})=\pi_{S}\left(\theta \mid \mathbf{x}^{obs}\right), \quad \forall \theta \in \Theta .
$$
\end{theorem}
This theorem is universally applicable to any scoring rule for which unbiased estimates can be obtained. In this context, we employ the kernel scoring rule \citep{pacchiardi2023generalized},
\begin{equation}
	S_k(P_{\theta}, \mathbf{x}^{obs}) = \E[k(\mathbf{x},\mathbf{x}')] - 2\cdot\E [k(\mathbf{x}, \mathbf{x}^{obs})],\quad  \mathbf{x} \perp \mathbf{x}' \sim P_{\theta},
 \label{eq:kernel_score}
\end{equation}
for which an unbiased estimate (w.r.t. $P_{\theta}$) can formally be expressed as:
\begin{equation}
\widehat{S}_k \left(\mathcal{X}^{\theta,m}, \mathbf{x}^{obs}\right)=\frac{1}{m(m-1)} \sum_{\substack{j, q=1 \\ q \neq j}}^m k\left(\mathbf{x}^{\theta,j}, \mathbf{x}^{\theta,q}\right)-\frac{2}{m} \sum_{j=1}^m k\left(\mathbf{x}^{\theta,j}, \mathbf{x}^{obs}\right),
\label{eq:unbiased_kernel_score}
\end{equation}
when $k(\cdot, \cdot)$ is a symmetric and positive-definite kernel.
The specific choice of the kernel function $k$ for our temporal haplotype frequencies data will be elaborated upon in subsequent sections of this manuscript. It is worth noting that many popular kernels are not directly applicable to our haplotype frequencies data due to their temporal nature.

Utilizing the scoring rule posterior, along with an unbiased estimate of the kernel scoring rule, we are able to draw asymptotic samples from the target posterior distribution using Markov Chain Monte Carlo methods. To sample from our scoring rule posterior $\pi_{S}\left(\theta \mid \mathbf{x}^{obs}\right)$, we implement the Metropolis-Hastings algorithms \citep{hastings1970monte} here, which aligns with the pseudo-marginal approach of \citep{Andrieu_2009, pacchiardi2023generalized}. The proposal distribution for the Metropolis-Hastings algorithm was chosen as a zero mean multivariate Gaussian distribution with a covariance matrix being the identity matrix multiplied by a scaling constant $c$. We notice that the proposal distribution is on $\mathbb{R}^n$, but in the examples below we would see that parameters would be in bounded regions (e.g. $[a, b]$) or on a simplex. Hence, we would transform the parameters to the real line and perform the MCMC algorithm above on the real line. The transformations used here are described in the Appendix~\ref{sec:trasnform_mcmc}. 

\subsection{Signature kernel score}
\label{sec:signatures}

In our prior discussion, we advocated for the use of the kernel scoring rule and provided an unbiased estimator in equation (\ref{eq:unbiased_kernel_score}), while we did not offer a specific kernel choice. In this subsection, we clarify our selection, the Signature Kernel, as introduced in \citep{kiraly2019kernels}. 
We first describe some preprocessing steps needed to convert our time-series observations on discrete time points to continuous paths and to extract the most information from them.
\paragraph{Pre-processing:}
The observed and simulated data in our setup (e.g., allele frequencies sampled at discrete time points in our study) do not have a continuous path; rather, observations are on discrete time points similar to any other real-world applications. Hence, first we use a piecewise linear interpolation to derive a continuous path, which is needed with the Signature Kernel. 
Finally, following \citep{salvi2021signature} we
lift these continuous paths in the space $\mathcal{Y}$ to an infinite-dimensional informative feature space by means of a
feature map (to do so, we employ the ``kernel trick'' \citep{hofmann2008kernel} by using a ``static'' kernel, in this case, a Radial Basis Function (RBF) kernel $\left(k(u, v) = \exp(-\gamma \| u - v \|^2)\right)$,
where \( \gamma > 0 \) is a hyperparameter) on data, and then we learn signature kernel scores on these
lifted paths. Feature maps like this ensure that the uplifted space is 
a reproducing kernel Hilbert space (RKHS) $\left(\mathcal{H},k \right)$ with a kernel $k$ on $\mathcal{Y}$.
The reproducing property tells us, 
$$k:\mathcal{Y}\times\mathcal{Y}\rightarrow \mathbb{R} \mbox{ equals } k(u,v)=\langle k_u, k_v \rangle_{\mathcal{H}}$$
where $\mathcal{Y} \ni u \mapsto k_u(\cdot):= k(u, \cdot) \in \mathcal{H}$ can be considered as the (canonical) feature map
induced by the kernel $k$. Now a path on the space $y\in \mathcal{P}_{\mathcal{Y}}$ would be lifted to a path on the Hilbert space $k_y \in \mathcal{P}_{\mathcal{H}}$, given as $t \mapsto k_y(t) := k_{y(t)}(\cdot) := k(y(t), \cdot)$. 
\paragraph{Signature kernel:} Next, we introduce a mapping from a sufficiently smooth path (continuous time) $h\in \mathcal{P}_{\mathcal{H}}$ defined on a Hilbert space $\mathcal{H}$
to features called signatures, which are defined by iterated path integrals.
The \textit{signature} $(\mathrm{Sig}(h))$ of a path is then defined as an infinite series of tensors that captures all possible iterated integrals, systematically including the path's multifaceted dynamics \citep{chevyrev2016primer}. Analogous to moments in statistics, which capture various characteristics of a distribution of vector-valued random variables, the signature of a path characterizes the high-dimensional temporal data. 
Further, the space of signatures $\left(Sig(h): h\in \mathcal{P}_{\mathcal{H}}\right)$ with an underlying Hilbert space $\mathcal{H}$, is a linear space and indeed a Hilbert space after restricting to a subspace and completion. Hence the inner product $\langle Sig(g),Sig(h)\rangle$ of the signature features of (sufficiently regular) paths $g,h \in \mathcal{P}_{\mathcal{H}}$, which is finite, can be used to define the \emph{signature kernel} between the paths on the Hilbert space. After extracting the signature features of this new path on the Hilbert space $\mathcal{H}$, our combined feature map becomes, 
$$\mathcal{P}_{\mathcal{Y}} \ni y \mapsto Sig(k_y).$$ 
Theorem 1 of \citep{kiraly2019kernels} guarantees that $Sig(k_y)$ captures all the information about $y \in {P}_{\mathcal{Y}}$ while looking at the path $y$ through the lens of kernel $k$. \emph{Signature kernel} between paths in $x, y \in \mathcal{P}_\mathcal{Y}$ can be defined as 
$$(x,y)\mapsto k^{\bigotimes}(x, y) = \langle Sig(k_x), Sig(k_y) \rangle.$$
Theorem 2 of \citep{kiraly2019kernels} provides a proof via construction that $k^{\bigotimes}(x, y)$ can be calculated efficiently when we can evaluate $k(x(s), y(t))$ for $s, t \in [0, T]$, and the ``static'' kernel $k : \mathcal{Y} \times \mathcal{Y} \rightarrow \mathbb{R}$.  
In practice for the calculation of this estimate, we utilize the Python package \textsf{sigkernel} from \citep{salvi2021signature}, which uses the fact that the signature kernel is the solution to a class of hyperbolic partial differential equations known as the Goursat PDE, which can be efficiently solved using numerical solvers.

By plugging in the Signature kernel $\left( k^{\bigotimes}(x, y)\right)$ 
as described above, into equations~\ref{eq:kernel_score} and \ref{eq:unbiased_kernel_score},  we obtain the Signature kernel score and an unbiased estimate of the same. 

\section{Inference of Parametric Population Genetic Models}
\label{sec:inf_selcof}

In this section, we offer an evaluation of generalized Bayesian likelihood-free inference with Signature kernel score (GBLFI-SigSR) by testing its performance through simulations across diverse scenarios consisting of different true selection coefficients and recombination rates under both the standard Wright-Fisher model (Section~\ref{subsection:WFD}) and the frequency-dependent selection model (Section~\ref{sec:FDS}).
We focus our investigations on allele frequencies defined by the underlying one-locus, two-locus, and three-locus genotype combinations.

\paragraph{Benchmark Methods} We compare our proposed methodology with the HMM-WFD method proposed in  \citep{he2020detecting} and a simpler frequentist method proposed in \citep{taus2017quantifying}, called LLS, applied to allele frequency data from each locus individually. The methodology proposed in HMM-WFD also provides us with an approximate posterior distribution similar to ours. In contrast, \citep{taus2017quantifying} proposes a straightforward approach, which utilizes the basic linear relationship between logit-transformed allele frequencies and the number of generations \citep{taus2017quantifying}. Specifically, for the haploid case and the diploid case, respectively, we have:
\begin{subequations}\label{eq:linear}
    \begin{align}
        \ln \left( \frac{a_t}{1 - a_t} \right) &= \ln \left( \frac{a_0}{1 - a_0} \right) + st, \label{eq:haploidlinear}\\
        \ln \left( \frac{a_t}{1 - a_t} \right) &= \ln \left( \frac{a_0}{1 - a_0} \right) + \frac{s}{2}t. \label{eq:diploidlinear}
    \end{align}
\end{subequations}
Here, $a_t$ is the expected allele frequency at the final time point, $a_0$ is the initial expected allele frequency, $s$ represents the selection coefficient, and $t$ is the number of generations elapsed. Solving these equations, they provide us with an estimate of the selection coefficients at each of the selected loci. 

Given $\mathbf{x}^{obs,r}$ as observed data, we apply our method to each of these data sets leading to $n$ samples $(\theta_{1,r},\ldots,\theta_{n,r})$ from the respective scoring rule posterior distribution $\pi_{S}\left(\theta \mid \mathbf{x}^{obs,r}\right)$, using a Metropolis-Hastings algorithm starting from the LLS estimates. For a fair comparison between two Bayesian methods providing approximate posteriors and a frequentist method providing only an estimate, we compute the approximate posterior mean from these posterior samples to provide a point estimate. Scott's method \citep{scott2015multivariate} has been used for bandwidth selection. It is important to note that the LLS method obtains a single estimate of the selection coefficient for each locus, whereas our proposed method computes the posterior modes for each locus jointly. We acknowledge that the comparison may not be fair since a single locus method will be biased if selection is also present at linked neighboring loci. Thus, it may not be so surprising that our proposed method outperforms the LLS method in our simulation experiments.

To use our methodology, we needed to choose the parameter $\omega$ in Equation~\ref{eq:SR_post}, $\gamma$ the hyperparameter for the radial basis function mentioned in Section~\ref{sec:signatures} and the scaling constant $c$ for the Metropolis-Hastings algorithm in Section~\ref{sec:sampling}. For each experiment, we chose them based on pilot runs to achieve the acceptance rate for Metropolis-Hastings algorithm between $0.25-0.4$ \citep{roberts2001optimal}. This resulted in $\gamma$ being chosen as $0.1$ for all experiments, $\omega$ being chosen as 1 for all experiments except for yeast data, where it was necessary to choose 5 to compensate for the availability of fewer time-steps (can be interpreted as information) in the data. These values and those chosen for $c$, for all experiments, are reported for each experiment below.  

As our performance measure, we employ the Root Mean Square Error (RMSE) of the estimates for the selection coefficients. 
\begin{definition}[Root Mean Square Error (RMSE)]
\begin{equation}\label{defRMSE}
RMSE(s^{*}_r,s^{true}) = \sqrt{\frac{1}{d} \sum_{i=1}^{d} (s^{*}_{r,i} - s^{true}_{i})^2}
\end{equation}
where $s^{*}_r$ represents the estimated selection coefficient from the $\mathbf{x}^{obs,r},$ $(1\le r\le N_r)$ observed data, obtained as the posterior mode or LLS estimate. $s^{true}$ is the true selection coefficient used to simulate $N_r=10$ observed data under the same experimental scenario, and $d$ stands for the number of selection coefficients we are estimating, e.g. d=2 for the two-locus standard Wright-Fisher model. 
\end{definition}

\subsection{Discrete time Wright-Fisher (WF) Model}
\label{sec:Inf_WF}
The goal of this work is to infer selection coefficients from temporal allele frequency data when more than one locus is selected. We therefore illustrate the performance of our methodology in comparison to existing benchmark methods for the two-locus and three-locus Wright-Fisher models. In principle, the proposed methodology can be extended to situations where an arbitrary number of selected loci is considered simultaneously. 

All subsequent simulation studies consider the allele frequencies from a founding population $t_0$ to $t_0 + K \Delta t$, where $K \in \{1,2,3,4,5,6,7,8,9,10\}$ and $\Delta t$ is 10, representing a 10-generation interval assuming the initial haplotype frequency at $t_0$ was known. For each experimental scenario, we simulate 
$N_r=10$ data sets $\mathbf{x}^{obs,r},$ $(1\le r\le N_r)$ under the same known parameter values. 
The performance of our method will be assessed based on these
ten replicates. In experimental evolution, often also multiple experimental replicates are available. In this case, posterior samples
might be aggregated across replicates.

\begin{figure}[htbp!]
    \centering
    \includegraphics[width=0.49\linewidth]{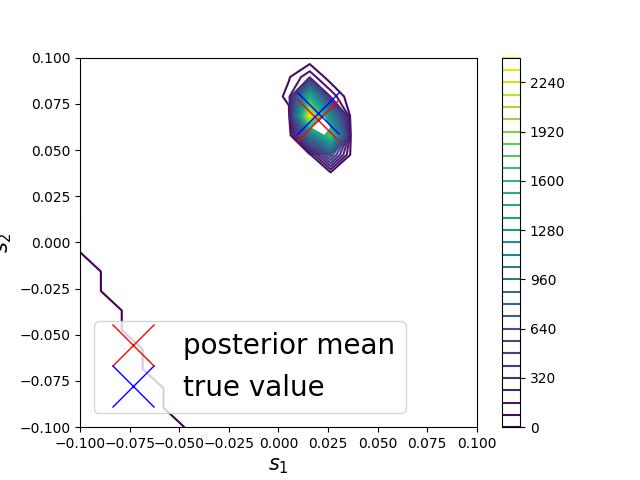}
    \caption{\textbf{Inference of selection coefficients for a two-locus scenario:} Posterior distribution of $(s_1, s_2)$ from a trajectory simulated with true selection coefficients [0.02, 0.07], recombination rate [1e-6], population size 5000, 100 generations, observed every 10 generations. Blue and red crosses: true value of selection coefficient and posterior mean ($[\hat{s}_1, \hat{s}_2]=[0.02  0.066]$) correspondingly.}
    \label{fig:2L_example}
\end{figure}

\paragraph{Two-locus Wright-Fisher model:}  With our simulated trajectories, we assumed initial haplotype frequencies to be known and given as $[0.1, 0.2, 0.3, 0.4]$. In Figure~\ref{fig:2L_example}, we illustrate the obtained posterior distribution and posterior mean of $(s_1, s_2)$ from a trajectory simulated with true selection coefficient [0.02, 0.07], showing a very good agreement. In this and the following experiments, we chose $\gamma=0.1$, and the Metropolis-Hastings algorithm with scaling coefficient $1e-4$ was run for 1000 steps with a burn-in of 200 steps. The comparison in Table~\ref{tab:2LWDF_simplified} incorporates five different values for the recombination rate $r_{AB}$, specifically 0.000001, 0.01, 0.1, 0.5, and 0. We examined 4 distinct pairs of selection coefficient values ($s_1,s_2$). For each pair, we conducted 10 repetitions ($N_r = 10$) under the same setup. For each experiment, we obtained 300 posterior samples from each of the methods. The results show that our method outperforms the HMM-WFD \citep{he2020detecting} and LLS \citep{taus2017quantifying} methods across the studied scenarios. By comparing the performance under various recombination levels and selection coefficients, we demonstrate the good performance of our proposed method in a wide range of evolutionary scenarios. 

 \begin{table}[htbp!]
    \centering
    \fontsize{8.5}{8.5}\selectfont
    \renewcommand{\arraystretch}{0.45}
    \setlength{\tabcolsep}{3pt}
\begin{tabular}{lcccc}
\toprule
Recombination rate & Selection strength & \multicolumn{3}{c}{RMSE} \\
\cmidrule(lr){3-5}
& & GBLFI-SigSR & HMM-WFD \citep{he2020detecting} & LLS \citep{taus2017quantifying} \\

$r_{AB}=0$ & s : (0.02, 0.02) & \textbf{0.0021} (0.0004) & 0.0181 (0.0125) & 0.0262 (0.0004)\\
& s : (0.02, 0.05) & \textbf{0.0025} (0.0007) & 0.0171 (0.0116) & 0.0490 (0.0010)\\
& s : (0.02, 0.07) & \textbf{0.0042} (0.0010) & 0.0254 (0.0106) &0.0654 (0.0020)\\
& s : (0.02, 0.09) & \textbf{0.0041} (0.0012) & 0.0192 (0.0127) &0.0811 (0.0012)\\

\midrule
$r_{AB}=1e-6$ & s : (0.02, 0.02) & \textbf{0.0021} (0.0004) & 0.0225 (0.0123) & 0.0262 (0.0004)\\
& s : (0.02, 0.05) & \textbf{0.0014} (0.0004) & 0.0229 (0.0125) & 0.0491 (0.0009)\\
& s : (0.02, 0.07) & \textbf{0.0027} (0.0009) & 0.0221 (0.0142) &0.0655 (0.0019)\\
& s : (0.02, 0.09) & \textbf{0.0041} (0.0033) & 0.0239 (0.0116) &0.0813 (0.0017)\\
\midrule
$r_{AB}=1e-2$ & s : (0.02, 0.02) & \textbf{0.0017} (0.0002) & 0.0181 (0.0078) & 0.0254 (0.0006)\\
& s : (0.02, 0.05) & \textbf{0.0008} (0.0007) & 0.0185 (0.0085) & 0.0474 (0.0009)\\
& s : (0.02, 0.07) & \textbf{0.0016} (0.0009) & 0.0205 (0.0059) &0.0634 (0.0013)\\
& s : (0.02, 0.09) & \textbf{0.0038} (0.0025) & 0.0223 (0.0118) &0.0802 (0.0020)\\
\midrule

$r_{AB}=0.1$ & s : (0.02, 0.02) & \textbf{0.0006} (0.0002) & 0.0134 (0.0065) &0.0241 (0.0005)\\
& s : (0.02, 0.05) & \textbf{0.0004} (0.0003) & 0.0115 (0.0067) &0.0459 (0.0003)\\
& s : (0.02, 0.07) & \textbf{0.0015} (0.0003) & 0.0128 (0.009) &0.0628 (0.0015)\\
& s : (0.02, 0.09) & \textbf{0.0033} (0.0010) & 0.0169 (0.0112) & 0.0796 (0.0016)\\
\midrule
$r_{AB}=0.5$ & s : (0.02, 0.02) & \textbf{0.0004} (0.0002) & 0.0172 (0.0039) &0.0239 (0.0003)\\
& s : (0.02, 0.05) & \textbf{0.0007} (0.0003) & 0.0141 (0.0066) &0.0458 (0.0005)\\
& s : (0.02, 0.07) & \textbf{0.0018} (0.0006) & 0.0163 (0.0133) & 0.0626 (0.0008)\\
& s : (0.02, 0.09) & \textbf{0.0035} (0.0033) & 0.0232 (0.0138) & 0.0782 (0.0014)\\
\bottomrule
\end{tabular}
    \caption{\textbf{RMSE of posterior modes for a two-locus standard WF model.} We considered our proposed method (GBLFI-SigSR), LLS \citep{taus2017quantifying},  and HMM-WFD \citep{he2020detecting} across different scenarios of selection strengths ($s$) and recombination rates ($r_{AB}$). Here we report the mean and (standard deviation) of the RMSE values over $N_r = 10$ repetitions for each experimental setup. We use the bold format to highlight the methods with the lowest mean RMSE. }
    \label{tab:2LWDF_simplified}
\end{table}

\begin{figure}[htbp!]
    \centering
    \begin{subfigure}[t]{0.45\textwidth}
    \centering
    \includegraphics[width=\linewidth]{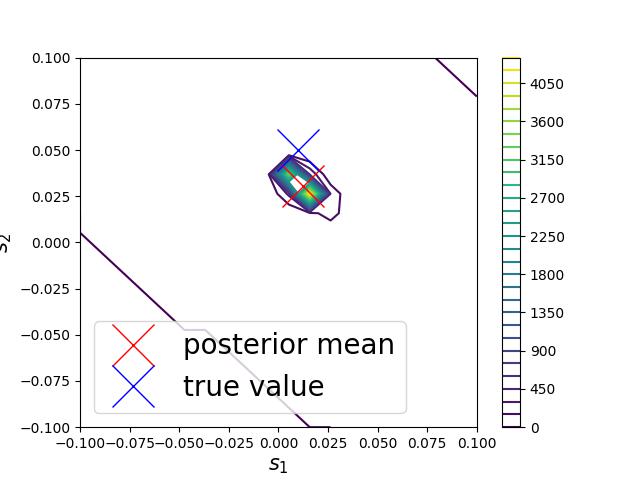} 
    \caption{Marginal posterior of $(s_1, s_2)$}
    \end{subfigure}
    ~
    \begin{subfigure}[t]{0.45\textwidth}
    \centering
    \includegraphics[width=\linewidth]{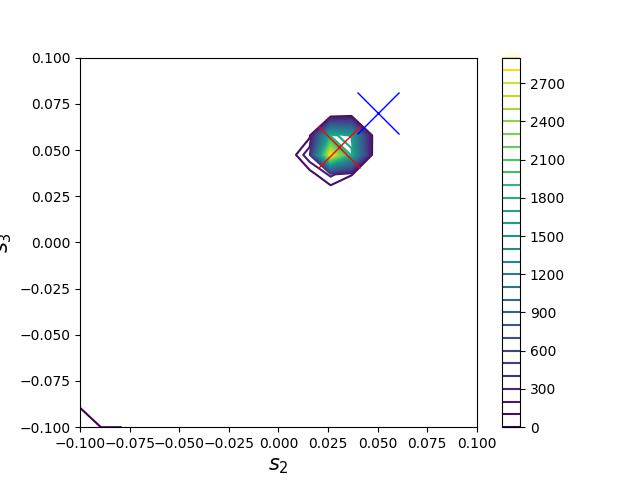} 
    \caption{Marginal posterior of $(s_2, s_3)$}
    \end{subfigure}
    ~
    \begin{subfigure}[t]{0.45\textwidth}
    \centering
    \includegraphics[width=\linewidth]{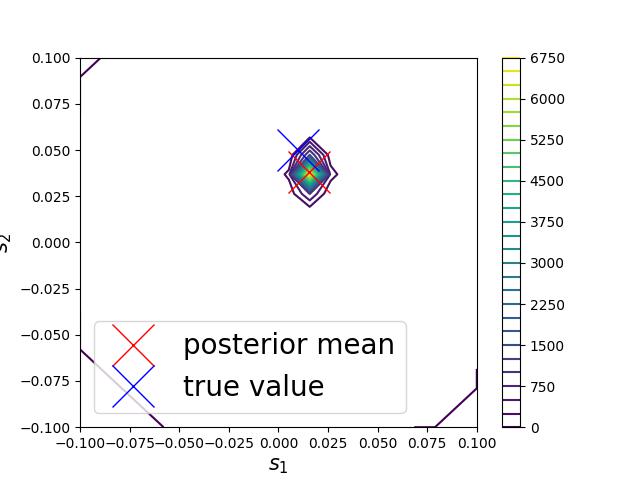} 
    \caption{Marginal posterior of $(s_1, s_2)$}
    \end{subfigure}
    ~
    \begin{subfigure}[t]{0.45\textwidth}
    \centering
    \includegraphics[width=\linewidth]{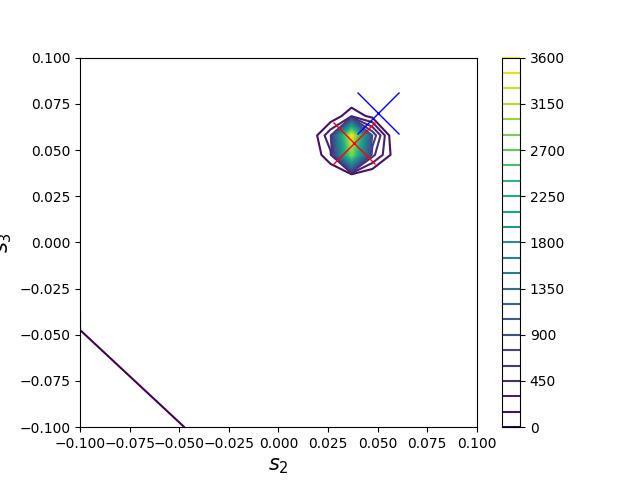} 
    \caption{Marginal posterior of $(s_2, s_3)$}
    \end{subfigure}
    \caption{\textbf{Inference of selection coefficient for a three-locus scenario:} Bivariate marginal posterior distributions of $(s_1, s_2, s_3)$ from a trajectory simulated with true selection coefficients [0.02, 0.03, 0.5], recombination rates [0.001,0.01] and [0.1,0.5] correspondingly in (a, b) and (c, d), population size [5000], total number of generations [100], observed at a generation interval [10]. The blue and red crosses denote the true value of the selection coefficient and the posterior means. Our estimates are $[\hat{s}_1, \hat{s}_2, \hat{s}_3]=[0.012, 0.03, 0.051]$ and $[\hat{s}_1, \hat{s}_2, \hat{s}_3]=[0.012, 0.03, 0.051]$ correspondingly in (a, b) and (c, d).}
    \label{fig:3L_posterior}
\end{figure}

\paragraph{Three-locus Wright-Fisher model:} In Table~\ref{tab:3ltable3}, we provide results when selection is simultaneously acting on three loci that generate eight haplotypes. The system also features two recombination rates between the adjacent loci. We consider both a low and a high recombination scenario with rates given as [0.001,0.01] and [0.1,0.5]. In Figure~\ref{fig:3L_posterior}, we illustrate the agreement of the posterior and its mean with the true parameter value for a single experiment. To simulate the trajectories, we have fixed the initial haplotype frequencies to 0.0278, 0.0556, 0.0833, 0.1111, 0.1389, 0.1667, 0.1944, and 0.2222, and the population size to 5000. Furthermore, the total number of generations has been 100, with frequencies observed every tenth generation. For all the experiments here, we chose $\gamma=0.1$.
The Metropolis-Hastings algorithm with scaling coefficient $1e-4$ was run for 1000 steps with a burn-in of 300 steps.

\begin{table}[htbp!]
        \centering
        \floatsetup[table]{captionskip=0pt} 
        \begin{tabular}[ht]{llllrr}
        \toprule
            &             & & GBLFI-SigSR & LLS \citep{taus2017quantifying} \\
        Experiments & Selection Coefficients & Recombination Pairs &           &           \\
\midrule
 & $s_1 = 0.01$ &  $r_{12} = 0, r_{23} = 1e-5 $ & \textbf{0.0145} (0.0033) & 0.0611 (0.0007)  \\
 & & $r_{12} = 1e-4, r_{23} = 1e-3$ & \textbf{0.0141} (0.0054) & 0.0614 (0.0009)  \\
Scenario 1 &$s_2 = 0.05$ & $r_{12} = 1e-3, r_{23} = 1e-2$ & \textbf{0.0188} (0.0060) & 0.0604 (0.0005) \\
 & &  $r_{12} = 1e-2, r_{23} = 1e-1$ & \textbf{0.0161} (0.0035) & 0.0606 (0.0007)  \\
& $s_3 = 0.07$ & $r_{12} = 1e-1, r_{23} = 5e-1$ & \textbf{0.0115} (0.0054) & 0.0600 (0.0004)  \\
 & &  &  & \\
 \midrule
 & $s_1 = 0.02$ &  $r_{12} = 0, r_{23} = 1e-5 $ & \textbf{0.0097} (0.0049) & 0.0436 (0.0007)  \\
 & & $r_{12} = 1e-4, r_{23} = 1e-3$ & \textbf{0.0072} (0.0034) & 0.0437 (0.0006)  \\
Scenario 2 &$s_2 = 0.03$ & $r_{12} = 1e-3, r_{23} = 1e-2$ & \textbf{0.0092} (0.0027) & 0.0434 (0.0003) \\
 & &  $r_{12} = 1e-2, r_{23} = 1e-1$ & \textbf{0.0066} (0.0053) & 0.0435 (0.0006)  \\
& $s_3 = 0.05$ & $r_{12} = 1e-1, r_{23} = 5e-1$ & \textbf{0.0057} (0.0025) & 0.0428 (0.0005)  \\
 & &  &  & \\
        \bottomrule
        \end{tabular}
        \caption{\textbf{RMSE of posterior modes for a three-locus standard Wright-Fisher model:} Results for our proposed method (GBLFI-SigSR) and LLS \citep{taus2017quantifying} for two scenarios with true selection coefficients $\{s_1,s_2,s_3\}$
        chosen \{0.01,0.05,0.07\} and \{0.02,0.03,0.05\} respectively. 
        Different pairs of recombination rates were investigated. Their
        values are displayed in the table.
        Here we report the mean and (standard deviation) of the RMSE values over $N_r = 10$ repetitions for each experimental setup. We use bold format to highlight the methods with a lower mean RMSE. }
        \label{tab:3ltable3}
    \end{table}

\subsection{Negative Frequency Dependent Selection (NFDS) Model }
\label{sec:Inf_NFDS}
\begin{figure}[htbp!]
    \centering
        \begin{subfigure}[t]{0.45\textwidth}
    \centering
    \includegraphics[width=\linewidth]{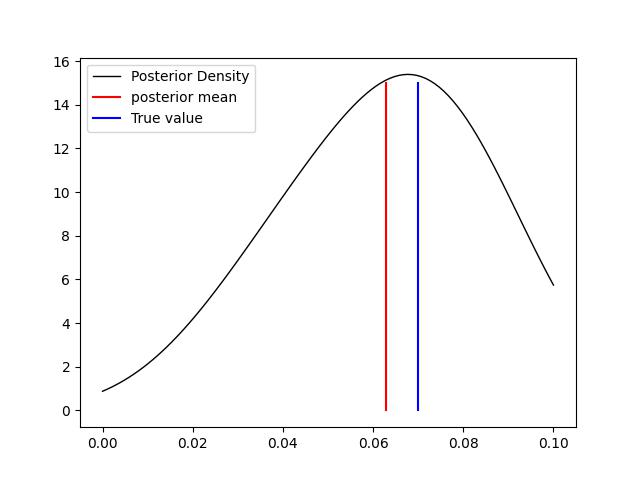} 
    \caption{One selected locus: Generalized posterior of $(s)$}
    \label{fig:1L_FDS}
    \end{subfigure}
     ~
    \begin{subfigure}[t]{0.45\textwidth}
    \centering
    \includegraphics[width=\linewidth]{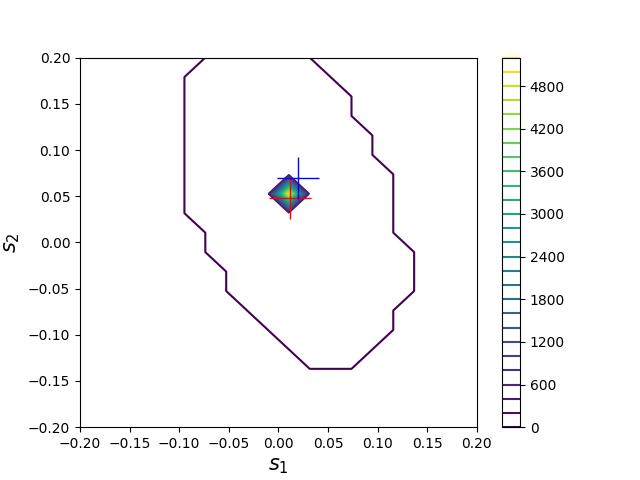} 
    \caption{Two selected loci: Generalized posterior of $(s_1, s_2)$}
    \label{fig:2L_FDS}
    \end{subfigure}
    \caption{\textbf{Inference of selection coefficients of Simulated Data from NFDS with one or two selected loci:}: \textbf{(a)} The posterior distributions of $s$ from a trajectory simulated with true selection coefficient $(s) = [ 0.07]$, recombination rate [1e-2], population size [5000], total number of generations [100], observed at generation interval [10]. The blue and red lines are the true value of the selection coefficient and the posterior mean $[\hat{s}] = [0.062]$as an estimate. \textbf{(b)} The joint posterior distributions of $(s_1, s_2)$ from a trajectory simulated with true selection coefficient $(s_1, s_2) = [0.02, 0.07]$, recombination rate [1e-2], population size [5000], total number of generations [100], observed at generation interval [10]. The blue and red crosses are the true value of the selection coefficient and the posterior mean 
    $[\hat{s}_1, \hat{s}_2] = [0.01, 0.075]$ as an estimate. }
    \label{fig:FDS}
\end{figure}
As a further scenario, we now infer selection coefficients involving either one or two loci under negative frequency-dependent selection. 
In \citep{SatoY}, the effects of genotype similarity on fitness has been used to infer selection.
As we are not aware of methods that use multi-locus temporal 
allele-frequency data for inference, we just illustrate the performance of our method GBLFI-SigSR by demonstrating the agreement between the inferred posterior and the posterior means with the true parameter values used to simulate the trajectories. In Figure~\ref{fig:FDS}, we consider both a single-locus $(s=0.07)$ and a two-locus scenario with $(s_1,s_2) = (0.02, 0.07)$. 
For both scenarios, we chose a population size of 5000 in an experiment running over 100 generations, observed at a generation interval of ten.
In the two-locus case, we chose a recombination rate of 0.01.  We have chosen $\gamma=0.01$, and the Metropolis-Hastings algorithm was run for 2000 steps with a burn-in of 500 steps with the scaling constants $1e-2$
(one locus) and $1e-3$ (two-loci). 

\subsection{Real Data: Yeast data for three-locus inference}
\label{sec:yeast_data}  
Yeast is a popular organism used in experimental evolution studies. A comprehensive review on techniques and challenges with this model organism can be found in \citep{burke2023embracing}.
A publication with publicly available data in this context is \citep{phillips2021}, where strategies have been explored to maximize standing genetic variation. In experimental evolution, designs that start with large variation should, in principle, be better suited to explore signals of selection in terms of allele frequency change. This is explored by \citep{phillips2021} by looking at
Manhattan plots obtained by applying Pearson's $\chi^2$-test,
as well as looking at haplotype frequency changes at candidate regions affected by selection. Here, we illustrate that our approach may be used to obtain a more detailed picture of the 
selective architecture, going beyond the analysis of the original publication. Data are available from this experiment for different numbers of starting haplotypes (4, 8, and 12) and two different methods of constructing founder populations (a simple K-method, and a more sophisticated S-method). The crossing scheme with eight starting haplotypes has been identified as the
most informative in \citep{phillips2021}. We therefore focus on this scheme which
can be analyzed in terms of our 
three-locus model, since three loci code for eight different haplotypes. 
The S-method of obtaining the founder population has been identified as the more reliable one.
The experiment involved 12 cycles of sexual reproduction. The total number of reproductions between cycles is not exactly known, but estimated to be between 15 and 20. 
Here, we follow the approach taken in \citep{chen2023haplotype} and set the total number of generations to $17.5*12=210$, and the effective population size to an estimated value of  $N_e=2000.$ 
The data set we used contains haplotype frequencies for regions of length 1kb (kilobase). These regions typically contain many SNPs, so that the identification of causal SNPs will not be possible. We therefore assume that putatively selected SNPs can either be proposed, or that three SNPs are chosen in order to uniquely identify each of the eight underlying haplotypes.
Here, we estimate selection coefficients for three representative SNPs that can 
explain the haplotype dynamics. Indeed, our fitness model translate selection coefficients for SNPs into haplotype frequencies. It should be noted that the data are available only at three time-points (including the starting time), which is still sufficient to get reasonable estimates of the underlying selection coefficients.
As an example, we consider a region on chromosome 11 between positions 615000 and 620000, since this region has already been identified as affected by selection in previous work by \citep{burke2014standing}.  We want to point out that the number of generations and population size are plausible values that have been chosen for illustrative purposes. For an actual biological application, a sensitivity analysis using a wider range of possible values should be carried out. Further algorithmic details of our application are: $\omega=5, \gamma=0.1$, scaling for MH is $1e-4$, chain length 1000, burn-in 300.

For the K-scheme, the LLS estimate ($[\hat{s}_1,\hat{s}_2,\hat{s}_3] = [-0.05,  0.024, -0.027]$) was used as the starting value of MCMC chain.
For the S-scheme, the starting values were taken as ($[\hat{s}_1,\hat{s}_2,\hat{s}_3] = [-0.0115,  0, -0.0011]$).

\begin{figure}[t!]
    \centering
    \begin{subfigure}[t]{0.49\textwidth}
    \centering
    \includegraphics[width=\linewidth]{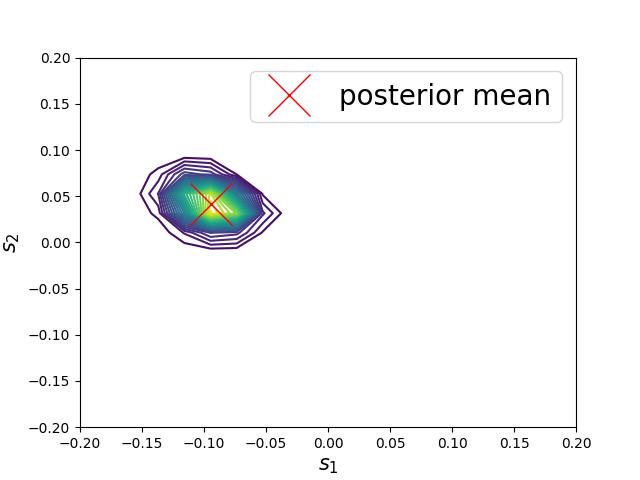} 
    \caption{Generalized posterior of $(s_1, s_2)$}
    \end{subfigure}
    ~
    \begin{subfigure}[t]{0.49\textwidth}
    \centering
    \includegraphics[width=\linewidth]{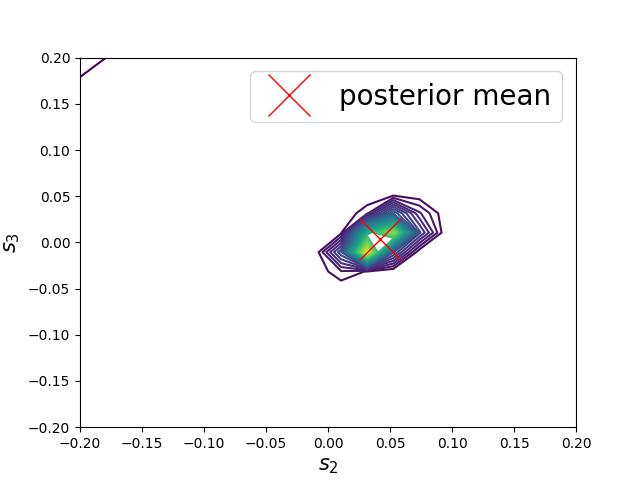} 
    \caption{Generalized posterior of $(s_2, s_3)$}
    \end{subfigure}
    \caption{\textbf{Inference of selection coefficient for Yeast Data from K-method:} The marginal posterior distributions (a,b) of $(s_1, s_2, s_3)$ from Yeast data assuming population size [2000], total number of generations [210], observed at generation interval [105, 210]. The black dots are posteriors samples, whereas the red cross is the posterior mean ($[\hat{s}_1,\hat{s}_2,\hat{s}_3] = [-0.094,  0.041,  0.004]$) as an estimate. }
    \label{fig:Yeast_K}
\end{figure}

\begin{figure}[t!]
    \centering
    \begin{subfigure}[t]{0.49\textwidth}
    \centering
    \includegraphics[width=\linewidth]{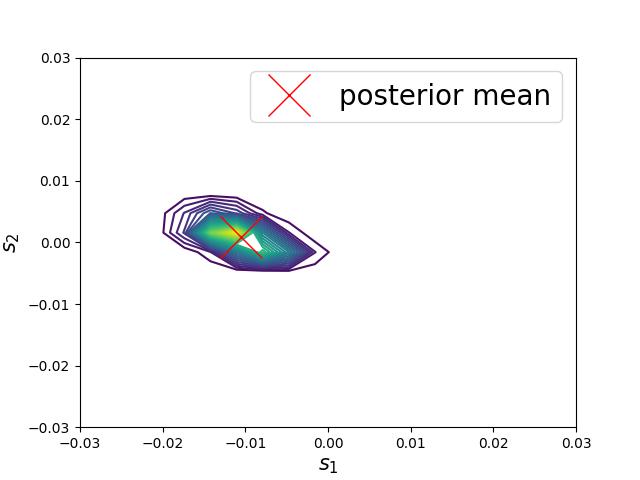} 
    \caption{Generalized posterior of $(s_1, s_2)$}
    \end{subfigure}
    ~
    \begin{subfigure}[t]{0.49\textwidth}
    \centering
    \includegraphics[width=\linewidth]{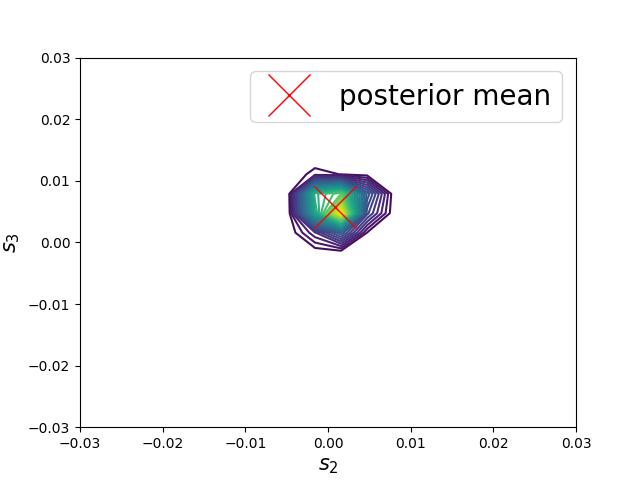} 
    \caption{Generalized posterior of $(s_2, s_3)$}
    \end{subfigure}
    \caption{\textbf{Inference of selection coefficient for Yeast Data from S-method:} The marginal posterior distributions (a,b) of $(s_1, s_2, s_3)$ from Yeast data assuming population size [2000], total number of generations [210], observed at generation interval [105, 210]. The black dots are posteriors samples, whereas the red cross is the posterior mean ($[\hat{s}_1,\hat{s}_2,\hat{s}_3] = [-0.0105,  0.0008,  0.0058]$) as an estimate. }
    \label{fig:Yeast_S}
\end{figure}

Our results from the three locus Wright-Fisher model are provided in Figures~\ref{fig:Yeast_K} and\ref{fig:Yeast_S}. They match the results obtained in Table 5 of \citep{phillips2021} but provide more details. 
Both our estimates and those in \citep{phillips2021} suggest that the dominant haplotype (Y12) is affected by negative selection. The selective signal is
also stronger for the K-scheme than for the S-scheme.
The inferred selection coefficients may be used to obtain fitness values for all underlying haplotypes.

\section{Joint Inference of selection coefficient and initial haplotype frequency}
\label{sec:joint_selcof_inthapfreq}
Until now, we have assumed that the initial haplotype frequency is known and focused on inferring the selection coefficients at the selected loci based on the allele frequency data. But in many experimentally observed datasets, we may only have access to the allele frequencies, while the initial haplotype frequencies are unknown. This applies, for instance, to the {\em Drosophila} dataset we will consider later in this section. In such a case, we propose to treat the initial haplotype frequencies as additional parameters and learn a generalized Bayesian posterior jointly on the initial haplotype frequencies and the selection coefficients.
For this purpose, we assume a Dirichlet distribution 
with all concentration parameters chosen $0.25$
as prior for the initial haplotype frequencies.  

\begin{figure}[htbp!]
    \centering
    \begin{subfigure}[t]{0.32\textwidth}
    \centering
    \includegraphics[width=\linewidth]{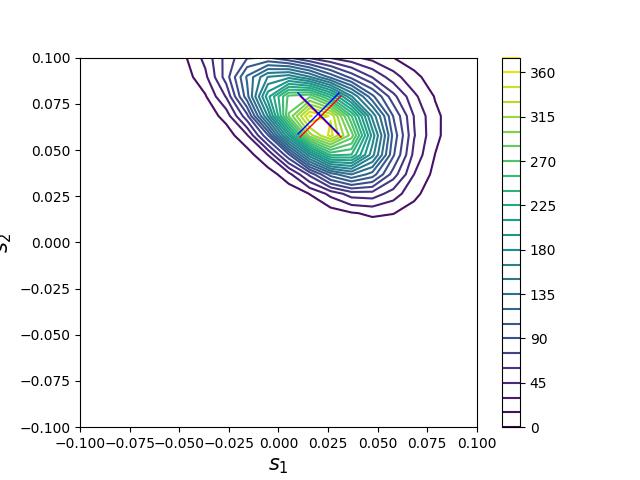} 
    \caption{Generalized posterior of $(s_1, s_2)$}
    \end{subfigure}
    ~
    \begin{subfigure}[t]{0.32\textwidth}
    \centering
    \includegraphics[width=\linewidth]{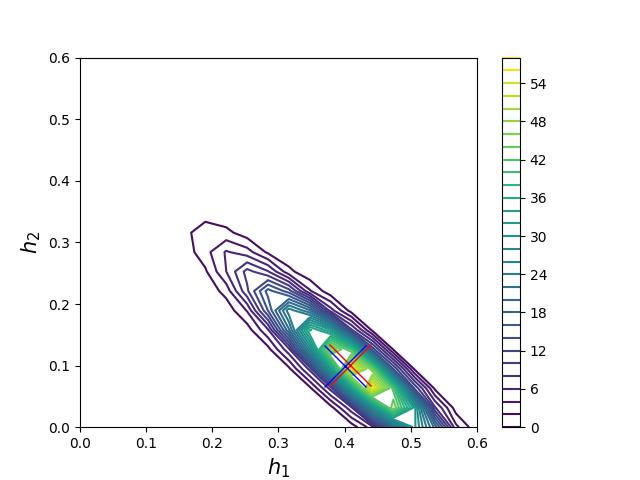} 
    \caption{Generalized posterior of $(h_1, h_2)$}
    \end{subfigure}
    ~
    \begin{subfigure}[t]{0.32\textwidth}
    \centering
    \includegraphics[width=\linewidth]{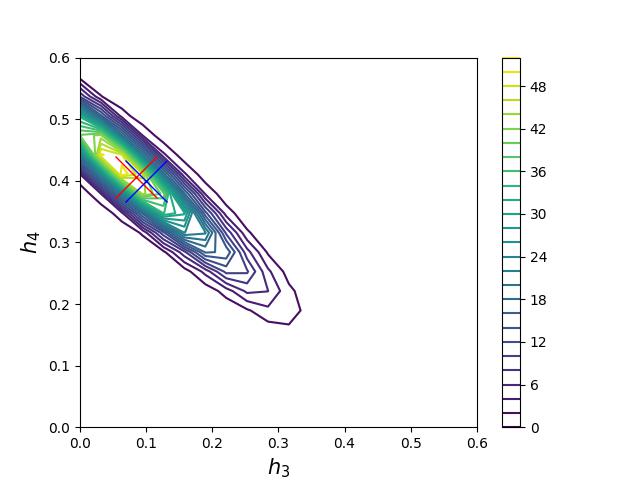} 
    \caption{Generalized posterior of $(h_3, h_4)$}
    \end{subfigure}

    \caption{\textbf{Inference of selection coefficients and initial haplotype frequencies for a two-locus scenario:} The marginal posterior distributions of $(s_1, s_2)$ and $(h_1, h_2, h_3, h_4)$ from a trajectory simulated with true selection coefficient $(s_1, s_2) = [0.02, 0.07]$, initial haplotype frequency $(h_1, h_2, h_3, h_4) = [0.4, 0.1, 0.1, 0.4]$, recombination rate [1e-6], population size [5000], total number of generation [100], observed at generation interval [10]. The blue and red crosses are the true value of the selection coefficient and the posterior mean as an estimate.}
    \label{fig:2L_joint_hap_sel}
\end{figure}
Before analysing a real data example, we first illustrate the performance of our inferential method in Figure~\ref{fig:2L_joint_hap_sel}. We  simulate again from a discrete time Wright-Fisher model with two selected loci, and selection coefficients $(s_1, s_2) = [0.02, 0.07]$, initial haplotype frequencies $(h_1, h_2, h_3, h_4) = [0.4, 0.1, 0.1, 0.4]$, recombination rate [1e-6], population size [5000], total number of generations [100], observed at a generation interval of ten. We report the inferred posterior, posterior mean (red cross, $[\hat{s}_1, \hat{s}_2]=[0.021, 0.068], ([\hat{h}_1, \hat{h}_2, \hat{h}_3, \hat{h}_4]=[0.408, 0.100, 0.085,   0.406]$) and true parameter values used for simulation, showing a very close agreement between them. In this case, we have chosen $\gamma=0.1$ and run Metropolis-Hastings algorithm for 100,000 steps with burn-in 10,000 and scaling constant $1e-3$.
\begin{figure}[htbp!]
    \centering
    \begin{subfigure}[t]{0.32\textwidth}
    \centering
    \includegraphics[width=\linewidth]{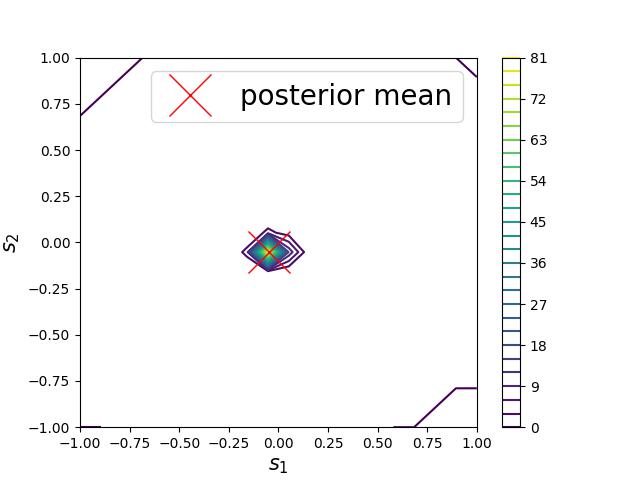} 
    \caption{Generalized posterior of $(s_1, s_2)$}
    \end{subfigure}
    ~
    \begin{subfigure}[t]{0.32\textwidth}
    \centering
    \includegraphics[width=\linewidth]{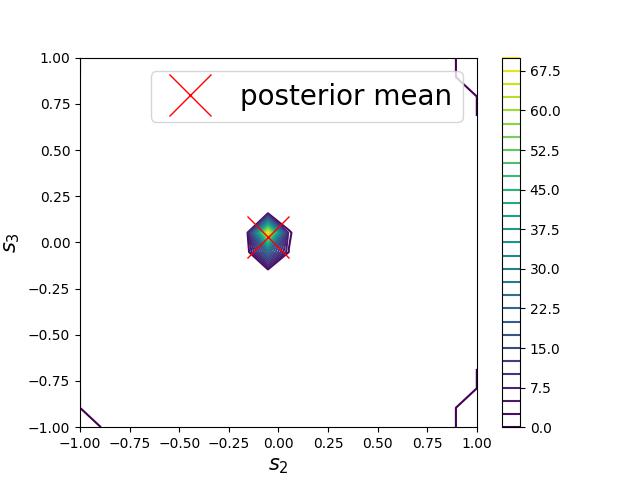} 
    \caption{Generalized posterior of $(s_2, s_3)$}
    \end{subfigure}
    ~
    \begin{subfigure}[t]{0.32\textwidth}
    \centering
    \includegraphics[width=\linewidth]{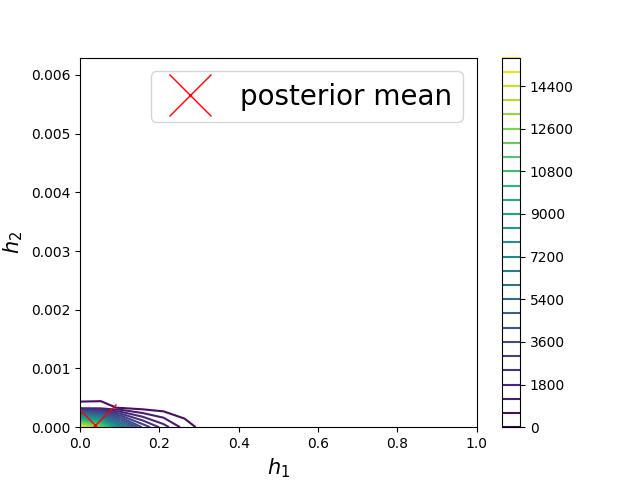} 
    \caption{Generalized posterior of $(h_1, h_2)$}
    \end{subfigure}
    ~
    \begin{subfigure}[t]{0.32\textwidth}
    \centering
    \includegraphics[width=\linewidth]{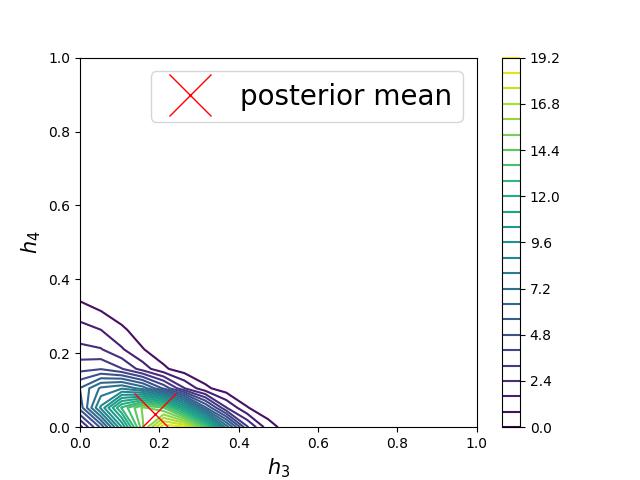} 
    \caption{Generalized posterior of $(h_3, h_4)$}
    \end{subfigure}
    ~
    \begin{subfigure}[t]{0.32\textwidth}
    \centering
    \includegraphics[width=\linewidth]{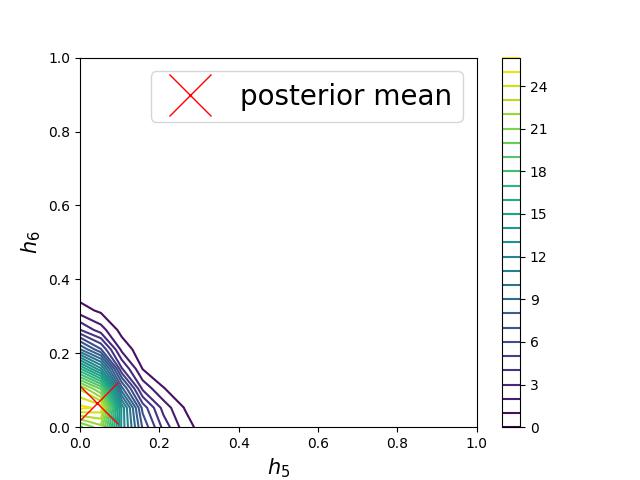} 
    \caption{Generalized posterior of $(h_5, h_6)$}
    \end{subfigure}
    ~
    \begin{subfigure}[t]{0.32\textwidth}
    \centering
    \includegraphics[width=\linewidth]{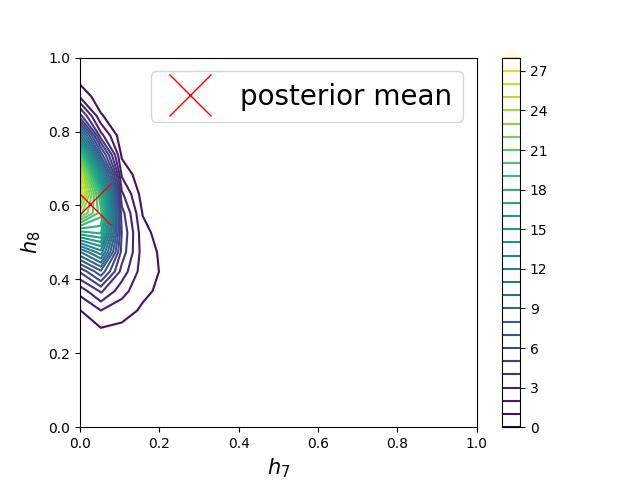} 
    \caption{Generalized posterior of $(h_7, h_8)$}
    \end{subfigure}
    \caption{\textbf{Inference of selection coefficients and initial haplotype frequency from three-locus {\em Drosophila} data:} The marginal posterior distributions of the selection coefficients and initial haplotype frequencies assuming recombination rates $[1.4e-3, 2.6e-4]$, population size [300], total number of generations [60], observed at generation interval [10]. The red cross represents the posterior mean as an estimate, ($[\hat{s}_1, \hat{s}_2, \hat{s}_3]= [-0.047, -0.051, 0.027]
    , [\hat{h}_1, \hat{h}_2, \hat{h}_3, \hat{h}_4, \hat{h}_5, \hat{h}_6, \hat{h}_7, \hat{h}_8] = 
    [0.036, 2.964e-05, 0.189, 0.034, 0.042, 0.066, 0.025, 0.604]
    $).}
    \label{fig:drosophila_analysis}
\end{figure}

\paragraph{Three-locus inference for an experiment on \textbf{\textit{ Drosophila simulans}}:}
Our objective is to evaluate the efficacy of our proposed approach in inferring selection coefficients, with a particular emphasis on actual three-locus data from an E\&R experiment, which included ten replications of a {\it Drosophila simulans} population subjected to a high-temperature regime over 60 generations \citep{barghi2019genetic}. Throughout the experiment, genomic DNA was systematically extracted from each replicated population every ten generations, employing the pool-seq method. The dataset utilized in our study contains more than 5 million SNPs. In addressing a three-locus case, we identified three SNP positions under selection and took into account the recombination between these loci. 
For this purpose, we used the recombination map provided by \citep{howie2019dna}. More specifically, the positions of these three loci are \{1314839,1363924,1372793\} on chromosome 2L, and the recombination in this window is 3 cM/Mb. Hence, we assumed the recombination rates as [1.4e-3, 2.6e-4]. 
The effective population size was set to 300, and the dominance coefficient was set to $h = 0.5$, adhering to the guidelines set forth by \citep{barata2023bait}. 

The data contained 10 replicates from 10 experiments, our inferred posterior is the distribution of the parameters given all 10 replicates. 
Assuming the initial haplotype frequencies are unknown, we infer the initial haplotype frequency and selection coefficients jointly. For inference, we chose $\gamma=0.1$ and ran Metropolis-Hastings algorithm for 100,000 steps with scaling constant $5e-3$ and burn-in 10,000. The inferred posterior distribution and posterior mean of the inferred selection coefficients and initial haplotype frequencies are displayed in Figure \ref{fig:drosophila_analysis}. The contour plot reveals a notable concentration of values, indicating a high degree of internal consistency within the posterior samples. This suggests that the estimates are fairly accurate in this example. We further notice that the mean of the estimates computed by the single locus LLS method of \citep{taus2017quantifying} for each of the 10 replicates is  [-0.005, -0.004, -0.003], which are closer to zero
compared to our method and third loci is negatively selected whereas our proposed method points to a positive selection at that loci. 

\section{Conclusion}
\label{section:conclusion}
Besides testing for selection, the estimation of selection coefficients is of considerable interest.  Researchers over the last two decades have developed methodologies capable of estimating selection coefficients, but the focus has been mostly on single loci.
A more recent study \citep{he2020detecting} has expanded the focus to consider scenarios with two linked loci, recognizing the importance of linkage dynamics. Our research extends these advances and takes them further. We consider scenarios with up to three linked loci 
under a Wright-Fisher model with selection, or alternatively a negative frequency-dependent selection model. We want to point out that an extension to more loci is possible in principle, as well as to other models of selection.
It would be interesting to explore up to which number of loci reasonably accurate estimates can be obtained. This may also depend on the underlying haplotype composition.

In this work, we have introduced a novel generalized likelihood-free Bayesian framework that employs signature-based kernels to approximate likelihoods and infer selection coefficients and initial haplotype frequencies for temporal population genetics models. This approach allows us to extract signatures from allele frequency trajectories, adeptly capturing the higher-order dynamics within the data and overcoming the intractability of traditional likelihood calculations.

Signature methods, novel tools stemming from rough path theory, have seen significant use in time series analysis. They offer a unique way to encapsulate the complex information embedded in time series data, capturing the underlying structure and intricate interactions effectively. Their broad application across various domains further confirms their efficacy and versatility. We apply signature kernels within a kernel scoring rule posterior in conjunction with a pseudo-marginal MCMC algorithm to obtain samples from an approximate generalized Bayesian posterior.

To conclude, our introduced signature-based likelihood approximation method showcases promising performance in quantifying natural selection in a multi-locus system. It extends existing methodology in that also three or potentially even more loci can be considered simultaneously.

\section*{Code Availability}
The corresponding Python codes can be found at https://github.com/statrita2004/MultiLocTempPopGenSigScore.
\section*{Acknowledgement}
We extend our sincere appreciation to Marta Pelizzola for her invaluable assistance in providing information about the {\it Drosophila simulans }data. R.D. is funded by EPSRC (grant nos. EP/V025899/1 and EP/T017112/1) and NERC (grant no. NE/T00973X/1). 
\bibliographystyle{siam}
\bibliography{ref}

\section{Appendix: Transformations used in MCMC}
\label{sec:trasnform_mcmc}

	We discuss here the transformations we apply in this work. 
    
    First, we will consider the case in which the support for the multivariate parameter $ x $ is defined by an intersection of element-wise inequalities, i.e. $ x \in \bigotimes_{i=1}^d (a_i,b_i) $, where $ a_i, b_i $ can take on the values $ \pm \infty $ as well. In this case, then, a transformation can be applied independently on each element of $ x $. We consider here the following transformations (which are also used in the \texttt{Stan} package \citep{carpenter2017stan}):
	
	\begin{itemize}
		\item When $ X \in [0, \infty)^d $, the transformation we use is $ y_i = \log (x_i) \in \R^d$. This corresponds to diagonal Jacobian with elements $ (J_{t}(x))^{-1}_{ii} = x_i $. 
		
		\item More generally, if $ x_i \in [a_i, +\infty) $  for $ |a_i| < \infty $, we can transform the data as $ y_i = \log (x_i - a_i) \in \R $, while if $ x_i \in (-\infty, b_i] $  for $ |b_i| < \infty $, we simply reverse the transformation: $ y_i = \log (b_i - x_i) \in \R $. These correspond to $ (J_{t}(x))^{-1}_{ii} = x_i - a_i$ and $ (J_{t}(x))^{-1}_{ii} = b_i - x_i$.
		
		\item Finally, if $ x_i \in (a_i,b_i) $ for $ |a_i|, |b_i| < \infty $, we can use the transformation defined as: $ y_i = t(x_i) = \logit \left(\frac{x_i - a_i}{b_i - a_i} \right) $ with inverse transformation $x_i = t^{-1} (y_i) = a + (b-a) \frac{e^{y_i}}{e^{y_i} + 1}$. This corresponds to $ (J_{t}(x))^{-1}_{ii} = \frac{(x_i - a_i)(b_i - x_i)}{b_i-a_i}$.	

	\end{itemize}

When the parameter $x$ lies in a $d$-dimensional simplex, s.t. $1\geq x_i\geq 0$ and $\sum_{i=1}^d x_i = 1$, then we consider a transformation \citep{porta2019sampling} extended by $r\in \mathbb{R}$. We transform $\mathbf{t} \in \mathbb{R}^{d}$ to $(r, x)$ as following, 
\begin{eqnarray*}
    r &=& t_1 + \ldots + t_d \\
    x_i &=& \frac{\exp{(t_i)}}{\exp{(t_1)} + \ldots + \exp{(t_d)}} \ \forall i=1,\ldots, d
\end{eqnarray*}

We can prove by induction on $d$, that the Jacobian determinant of the transformation would be,

\begin{equation}
|\mbox{det}\frac{d(r, x)}{dt}| = (d+1)\prod_{i=1}^d x_i \equiv (d+1)\exp{(r)} \left[\sum_i\exp{(t_i)}\right]^{-(d+1)}.
\end{equation}
The prior on $r$ can be arbitrary, which here we choose to be standard Normal distribution. 

\end{document}